\documentstyle[12pt,titlepage]{article}
\begin{document}
\begin{titlepage}
\parindent 0pt
{\Large\bf
Critical phenomena of the hard-sphere lattice gas
on the simple cubic lattice
}

\bigskip

{\large\bf Atsushi Yamagata}

{\it
Department of Physics, Tokyo Institute of Technology,
Oh-okayama, Meguro-ku, Tokyo 152, Japan

and

Department of Physics, Tokyo Metropolitan University,
Minami-ohsawa 1-1, Hachioji-shi, Tokyo 192-03, Japan
}

\bigskip

\begin{description}
\item[Running title]
Critical phenomena of hard-sphere lattice gas

\item[Keywords]
Hard-sphere lattice gas, Critical phenomena, Monte Carlo method

\item[PACS classification codes]
02.70.Lq, 64.60.Cn, 68.35.Rh

\end{description}

\bigskip

{\bf Abstract}

We study the critical phenomena of the hard-sphere lattice gas
on the simple cubic lattice with nearest neighbour exclusion
by the Monte Carlo method.
We get the critical exponents,
$\beta / \nu$ = 0.313(9) and
$\gamma / \nu$ = 2.37(2),
where $\beta$ is the critical exponent for the staggered density,
$\gamma$ for the staggered compressibility,
and $\nu$ for the correlation length.
\end{titlepage}

%%%%%%%%%%%%%%%%%%%%
%%% INTRODUCTION %%%
%%%%%%%%%%%%%%%%%%%%
\section{Introduction}
\label{sec:intro}
Atoms of a hard-sphere lattice gas occupy sites of a lattice and
interact with infinite repulsion of nearest neighbour pairs.
The grand partition function is
\begin{equation}
{\mit \Xi_{V}(z)}
=
\sum_{N} z^{N}\,Z_{V}(N),
\label{eqn:gpf}
\end{equation}
where $z$ is an activity and
$Z_{V}(N)$ is the number of configurations in which
there are $N$ atoms in the lattice of $V$ sites.
At $z=+\infty$ a ground state configuration is that the atoms occupy
all the sites of one sublattice and the other is vacant.
There is no atom at $z=0$.
A continuous phase transition occurs at a critical activity.
There are many studies of the system by various methods:
series expansions
\cite{Domb58,Temperley59,Burley60,GauntFisher65,Gaunt67},
finite-size scaling and transfer matrix
\cite{KamieniarzBlote93},
Bethe and ring approximations
\cite{Burley61b},
transfer matrix
\cite{Runnels65,RunnelsCombs66,ReeChesnut66},
corner transfer matrix and series expansions
\cite{Baxteretal80},
exact calculations
\cite{Baxter80},
and Monte Carlo simulations
\cite{Meirovitch83,Yamagata95}.

The critical activity is obtained on various lattice.
The square lattice:
3.80(2)
\cite{GauntFisher65},
3.80(4)
\cite{RunnelsCombs66},
3.7966(3)
\cite{ReeChesnut66},
3.7962(1)
\cite{Baxteretal80},
and
3.796255174(3)~\cite{BloteWu90}.
The triangular lattice:
11.12(11)
\cite{RunnelsCombs66},
11.05(15)
\cite{Gaunt67},
and
$\frac{1}{2}(11+5\sqrt{5}) = 11.090 \ldots$
\cite{Baxter80}.
The simple cubic lattice:
1.09(7)
\cite{Gaunt67}
and
1.0588(3)
\cite{Yamagata95}.
The body-centred cubic lattice:
0.77(5)
\cite{Gaunt67}
and
0.73(2)
\cite{Landau77,Racz80}.
The critical exponents on the square lattice are as follows.
The thermal exponent, $y_t$, is 1.000000(1),
the magnetic exponent, $y_h$, is 1.875000(1)
\cite{KamieniarzBlote93},
$\beta/\nu$ = 0.125(5),
and
$\gamma/\nu$ = 1.74(2)
\cite{Meirovitch83}
where $\beta$, $\gamma$, and $\nu$ are the critical exponents
for the order parameter, the staggered compressibility, and
the correlation length, respectively.
There are exact values on the triangular lattice
\cite{Baxter80,BaxterPearce82}:
$\alpha=1/3$ for the compressibility,
$\beta=1/9$,
$\mu=5/6$ for the interfacial tension, and
$\nu=5/6$.
The value, 0.61071(2)
\cite{KamieniarzBloteIsing93},
of the fourth-order cumulant
\cite{Binder81}
of the order parameter on the square lattice agrees with those
of the Ising model,
0.611(1)
\cite{Bruce85},
0.6103(7)
\cite{NicolaidesBruce88},
0.611(1),
and
0.610690(1)
\cite{KamieniarzBloteIsing93}.
These results support that the hard-sphere lattice gas on the square
lattice belongs to the universality class of the two-dimensional
Ising model and that on the triangular lattice does to one of the
three-state Potts model in two dimensions~\cite{Wu82}.
There is no result of critical exponents in three-dimensional
hard-sphere lattice gas.

We carry out Monte Carlo simulations of the hard-sphere lattice gas
on the simple cubic lattice.
Using the finite-size scaling,
we estimate the critical exponents.
In the next section we define physical quantities measured.
In section~\ref{sec:mcr} we describe procedure of analyses and
present Monte Carlo results.
A summary is given in section~\ref{sec:sum}.

%%%%%%%%%%%%%%%%%%%%%%%%%%%%%%%
%%% MONTE CARLO SIMULATIONS %%%
%%%%%%%%%%%%%%%%%%%%%%%%%%%%%%%
\section{Monte Carlo simulations}
We use the Metropolis Monte Carlo technique
\cite{Binder79,BinderStauffer87}
to simulate the hard-sphere lattice gas~(\ref{eqn:gpf})
on the simple cubic lattice
of $V$ sites, where $V$ = $L \times L \times L$
($L$ = $2 \times n$, $n$ = 2, 3, $\ldots$ , 30),
under fully periodic boundary conditions.
According to Meirovitch
\cite{Meirovitch83},
we adopt the grand canonical ensemble.
The algorithm is described in the references
\cite{Meirovitch83,Yamagata95}.

We start each simulation with a ground state configuration
at the critical activity,
$z_{\rm c} = 1.0588$
\cite{Yamagata95}.
The pseudorandom numbers are generated by the Tausworthe method
\cite{ItoKanada88,ItoKanada90}.
We measure physical quantities
over $1.2 \times 10^{6}$ Monte Carlo steps per site (MCS/site)
for $L \ge 44$ and $10^{6}$ MCS/site for $L \le 42$
after discarding $5 \times 10^{4}$ MCS/site to attain equilibrium.
We have checked that simulations from the ground state configuration
and no atom one gave consistent results.
Each run is divided into $M$ blocks.
Let us the average of a physical quantity, $O$, in each block
$\langle O \rangle_{i}$; $i$ = $1, 2, \ldots , M$.
The expectation value is
\[
\overline{\langle O \rangle}
=
\frac{1}{M}\,\sum_{i=1}^{M} \langle O \rangle_{i}.
\]
The standard deviation is
\[
{\mit \Delta} \langle O \rangle
=
\left(
\overline{\langle O \rangle^{2}} - \overline{\langle O \rangle}^{2}
\right)^{1/2}/\sqrt{M-1}.
\]
We have taken $M=12$ for $L \ge 44$ and $M=10$ for $L \le 42$.

Let us define an order parameter by
\[
R
=
2\,(N_{\rm A}-N_{\rm B})/V
\]
where $N_{\rm A}$ ($N_{\rm B}$) is the number of the atoms
in the A (B)-sublattice and $N = N_{\rm A} + N_{\rm B}$.
We measure the staggered densities,
\begin{equation}
m^{\dagger}
=
\overline{
\langle |R| \rangle
}
\label{eqn:stagdens}
\end{equation}
and
\begin{equation}
m^{\dagger \prime}
=
\overline{
\langle R^{2} \rangle^{1/2}
},
\label{eqn:rmsdensi}
\end{equation}
the staggered compressibilities,
\begin{equation}
\chi^{\dagger}
=
V
\overline{
\left(
\langle R^{2} \rangle - \langle |R| \rangle^{2}
\right)
}
/ 4
\label{eqn:stagcomp}
\end{equation}
and
\begin{equation}
\chi^{\dagger \prime}
=
V
\overline{
\langle R^{2} \rangle
}
/ 4,
\label{eqn:rmscompr}
\end{equation}
and the fourth-order cumulant of $R$ \cite{Binder81},
\begin{equation}
U
=
1-\frac{1}{3}\,
\overline{
\langle R^{4} \rangle / \langle R^{2} \rangle^{2}
}.
\label{eqn:stagcumu}
\end{equation}

%%%%%%%%%%%%%%%%%%%%%%%%%%%
%%% MONTE CARLO RESULTS %%%
%%%%%%%%%%%%%%%%%%%%%%%%%%%
\section{Monte Carlo results}
\label{sec:mcr}
We estimate a critical exponent and an amplitude
by using the finite-size scaling
\cite{Fisher70,Barber83,Privman90}.
For a physical quantitiy, $O$,
we use the nonlinear chi-square fitting
\cite{Pressetal92}
with a function of $L$,
\[
O(L)=A\,L^{p},
\]
where $p$ and $A$ are fitting parameters.
$p=-\beta/\nu$ when $O$ = $m^{\dagger}$ or $m^{\dagger \prime}$.
$p=\gamma/\nu$ when $O$ = $\chi^{\dagger}$ or $\chi^{\dagger \prime}$.
We calculate the chi-square per degrees of freedom,
$\chi^{2}/{\rm DOF}$,
and the goodness of fit, $Q$,
\cite{Pressetal92}
for the data set of the sizes
$L = L_{\rm min}, L_{\rm min}+2, \ldots , L_{\rm max}-2$, and
$L_{\rm max}$.
The values of $L_{\rm min}$ and $L_{\rm max}$ are selected so that
the difference between $\chi^{2}/{\rm DOF}$ and 1 is the smallest.

We get the results as follows.
For $m^{\dagger}$ defined by (\ref{eqn:stagdens}),
$\beta/\nu$ = 0.28(1),
$A$ = 0.39(2),
$\chi^{2}/{\rm DOF} = 1.21$, and
$Q = 0.27$.
For $m^{\dagger \prime}$ defined by (\ref{eqn:rmsdensi}),
$\beta/\nu$ = 0.313(9),
$A$ = 0.48(2),
$\chi^{2}/{\rm DOF} = 1.22$, and
$Q = 0.26$.
For $\chi^{\dagger}$ defined by (\ref{eqn:stagcomp}),
$\gamma/\nu$ = 1.96(2),
$A$ = 0.040(3),
$\chi^{2}/{\rm DOF} = 0.98$, and
$Q = 0.47$.
For $\chi^{\dagger \prime}$ defined by (\ref{eqn:rmscompr}),
$\gamma/\nu$ = 2.37(2),
$A$ = 0.058(4),
$\chi^{2}/{\rm DOF} = 1.21$, and
$Q = 0.28$.
$L_{\rm min} = 36 $ and $L_{\rm max} = 60$
for $m^{\dagger}$, $m^{\dagger \prime}$, and $\chi^{\dagger \prime}$.
$L_{\rm min} = 30 $ and $L_{\rm max} = 54$
for $\chi^{\dagger}$.

The values of the critical exponents of $m^{\dagger \prime}$ and
$\chi^{\dagger \prime}$ satisfy the scaling relation,
$2 \beta + \gamma = d\,\nu$
where $d$ is the dimensions, within errors.
Thus we adopt them as the reliable estimates.
Figure~1 shows the size dependence of $m^{\dagger \prime}$
at $z = z_{\rm c}$.
The solid line indicates $0.48 L^{-0.313}$.
Figure~2 shows the size dependence of $\chi^{\dagger \prime}$
at $z = z_{\rm c}$.
The solid line indicates $0.058 L^{2.37}$.
Figure~3 shows the size dependence of $U$
defined by (\ref{eqn:stagcumu}) at $z = z_{\rm c}$.
We cannot extract a reliable estimate in $L = +\infty$
since we do not know the correction to scaling of this system.
It will have a value between 0.54 and 2/3.

%%%%%%%%%%%%%%%
%%% SUMMARY %%%
%%%%%%%%%%%%%%%
\section{Summary}
\label{sec:sum}
We carry out the Monte Carlo simulations of the hard-sphere lattice
gas on the simple cubic lattice with nearest neighbour exclusion
under fully periodic boundary conditions.
Using the finite-size scaling,
we obtain the critical exponents,
$\beta/\nu$ = 0.313(9) and $\gamma/\nu$ = 2.37(2).
They satisfy the scaling relation, $2 \beta + \gamma = d \nu$.

The critical exponents of the three-dimensional Ising model
have been estimated,
$\beta/\nu$ = 0.518(7) and $\gamma/\nu$ = 1.9828(57)
\cite{FerrenbergLandau91}.
These values do not agree with
our results of the hard-sphere lattice gas.
The value of the fourth-order cumulant of the Ising model
on the simple cubic lattice under fully periodic boudary conditions
is as follows.
0.44(2)
\cite{Binder81},
0.4677(23)
\cite{LaiMon89},
and
0.4637(13)
\cite{BloteKamieniarz93}.
We see that these values are smaller than those of the hard-sphere
lattice gas as seen in figure~3.
While there is evidence that the hard-sphere lattice gas
on the square lattice
belongs to the Ising universality class in two dimensions
as described in section~\ref{sec:intro},
it does not seem that the hard-sphere lattice gas
on the simple cubic lattice falls into
the universality class of the three-dimensional Ising model.

%%%%%%%%%%%%%%%%%%%%%%%
%%% ACKNOWLEDGMENTS %%%
%%%%%%%%%%%%%%%%%%%%%%%
\section*{Acknowledgements}
We have carried out the simulations on the HITAC S-3600 computer
under the Institute of Statistical Mathematics Cooperative Research
Program (94-ISM$\cdot$CRP-43 and 95-ISM$\cdot$CRP-37) and
on two personal computers with the 486DX2/66MHz CPU and
the Linux operating system
(SLS 1.03 + JE 0.9.3, Slackware 2.0.0 + JE 0.9.5).
This study was supported by a Grant-in-Aid for Scientific Research
from the Ministry of Education, Science and Culture, Japan.

%%%%%%%%%%%%%%%%%%
%%% REFERENCES %%%
%%%%%%%%%%%%%%%%%%

%%%%%%%%%%%%%%%%%%%%%%%
%%% FIGURE CAPTIONS %%%
%%%%%%%%%%%%%%%%%%%%%%%
\clearpage
\section*{Figure captions}
\begin{description}
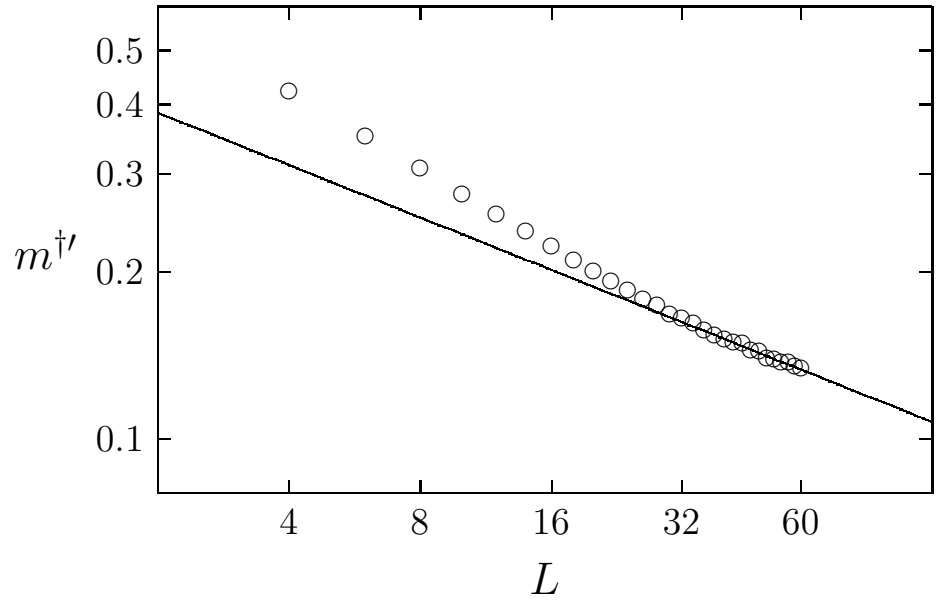
\item[Figure 1] Size dependence of the staggered density,
$m^{\dagger \prime}$, defined by (\ref{eqn:rmsdensi})
at $z = z_{\rm c}$.
The solid line indicates $0.48 L^{-0.313}$.
Errors are less than the symbol size.

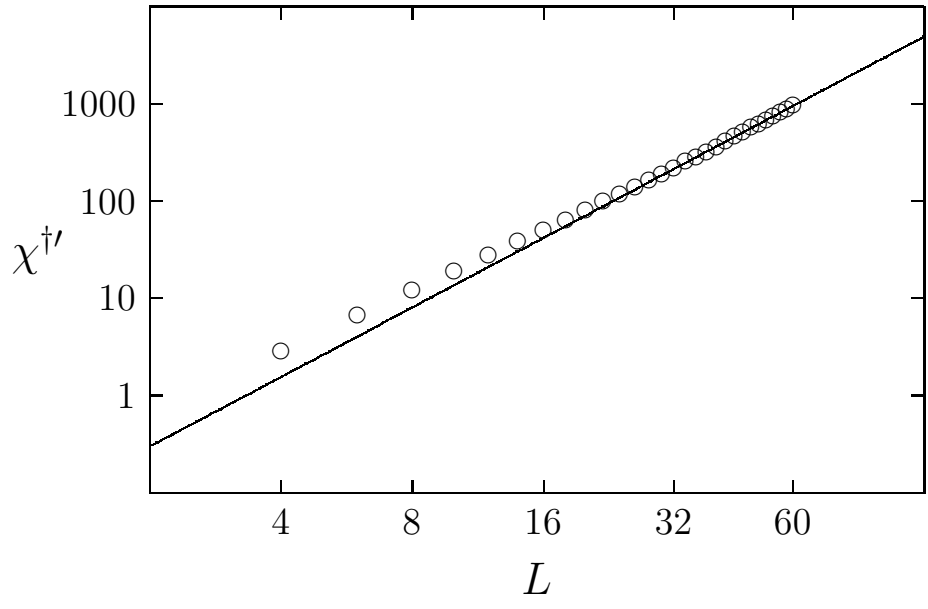
\item[Figure 2] Size dependence of the staggered compressibility,
$\chi^{\dagger \prime}$, defined by~(\ref{eqn:rmscompr})
at $z = z_{\rm c}$.
The solid line indicates $0.058 L^{2.37}$.
Errors are less than the symbol size.

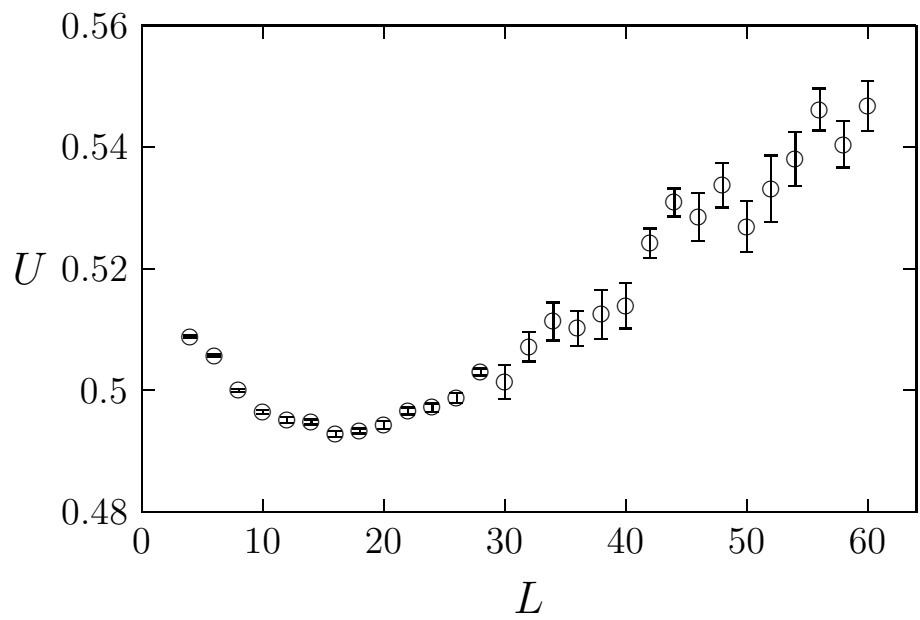
\item[Figure 3] Size dependence of the fourth-order cumulant,
$U$, defined by~(\ref{eqn:stagcumu})
at $z = z_{\rm c}$.
\end{description}

\clearpage
\begin{figure}
\begin{center}
% GNUPLOT: LaTeX picture
\setlength{\unitlength}{0.240900pt}
\ifx\plotpoint\undefined\newsavebox{\plotpoint}\fi
\sbox{\plotpoint}{\rule[-0.200pt]{0.400pt}{0.400pt}}%
\begin{picture}(1500,900)(0,0)
\font\gnuplot=cmr10 at 10pt
\gnuplot
\sbox{\plotpoint}{\rule[-0.200pt]{0.400pt}{0.400pt}}%
\put(220.0,198.0){\rule[-0.200pt]{4.818pt}{0.400pt}}
\put(198,198){\makebox(0,0)[r]{{\large $0.1$}}}
\put(1416.0,198.0){\rule[-0.200pt]{4.818pt}{0.400pt}}
\put(220.0,460.0){\rule[-0.200pt]{4.818pt}{0.400pt}}
\put(198,460){\makebox(0,0)[r]{{\large $0.2$}}}
\put(1416.0,460.0){\rule[-0.200pt]{4.818pt}{0.400pt}}
\put(220.0,614.0){\rule[-0.200pt]{4.818pt}{0.400pt}}
\put(198,614){\makebox(0,0)[r]{{\large $0.3$}}}
\put(1416.0,614.0){\rule[-0.200pt]{4.818pt}{0.400pt}}
\put(220.0,723.0){\rule[-0.200pt]{4.818pt}{0.400pt}}
\put(198,723){\makebox(0,0)[r]{{\large $0.4$}}}
\put(1416.0,723.0){\rule[-0.200pt]{4.818pt}{0.400pt}}
\put(220.0,808.0){\rule[-0.200pt]{4.818pt}{0.400pt}}
\put(198,808){\makebox(0,0)[r]{{\large $0.5$}}}
\put(1416.0,808.0){\rule[-0.200pt]{4.818pt}{0.400pt}}
\put(426.0,113.0){\rule[-0.200pt]{0.400pt}{4.818pt}}
\put(426,68){\makebox(0,0){{\large $4$}}}
\put(426.0,857.0){\rule[-0.200pt]{0.400pt}{4.818pt}}
\put(632.0,113.0){\rule[-0.200pt]{0.400pt}{4.818pt}}
\put(632,68){\makebox(0,0){{\large $8$}}}
\put(632.0,857.0){\rule[-0.200pt]{0.400pt}{4.818pt}}
\put(838.0,113.0){\rule[-0.200pt]{0.400pt}{4.818pt}}
\put(838,68){\makebox(0,0){{\large $16$}}}
\put(838.0,857.0){\rule[-0.200pt]{0.400pt}{4.818pt}}
\put(1043.0,113.0){\rule[-0.200pt]{0.400pt}{4.818pt}}
\put(1043,68){\makebox(0,0){{\large $32$}}}
\put(1043.0,857.0){\rule[-0.200pt]{0.400pt}{4.818pt}}
\put(1230.0,113.0){\rule[-0.200pt]{0.400pt}{4.818pt}}
\put(1230,68){\makebox(0,0){{\large $60$}}}
\put(1230.0,857.0){\rule[-0.200pt]{0.400pt}{4.818pt}}
\put(220.0,113.0){\rule[-0.200pt]{292.934pt}{0.400pt}}
\put(1436.0,113.0){\rule[-0.200pt]{0.400pt}{184.048pt}}
\put(220.0,877.0){\rule[-0.200pt]{292.934pt}{0.400pt}}
\put(45,495){\makebox(0,0){{\Large $m^{\dagger \prime}$}}}
\put(828,-22){\makebox(0,0){{\Large $L$}}}
\put(818,-412){\makebox(0,0){{\large Figure 1}}}
\put(818,-521){\makebox(0,0){{\large Atsushi Yamagata}}}
\put(220.0,113.0){\rule[-0.200pt]{0.400pt}{184.048pt}}
\put(426,744){\circle{24}}
\put(546,674){\circle{24}}
\put(632,623){\circle{24}}
\put(698,583){\circle{24}}
\put(752,551){\circle{24}}
\put(798,524){\circle{24}}
\put(838,500){\circle{24}}
\put(873,479){\circle{24}}
\put(904,462){\circle{24}}
\put(932,446){\circle{24}}
\put(958,431){\circle{24}}
\put(982,418){\circle{24}}
\put(1004,408){\circle{24}}
\put(1024,394){\circle{24}}
\put(1043,387){\circle{24}}
\put(1061,379){\circle{24}}
\put(1078,368){\circle{24}}
\put(1094,361){\circle{24}}
\put(1110,354){\circle{24}}
\put(1124,350){\circle{24}}
\put(1138,349){\circle{24}}
\put(1151,338){\circle{24}}
\put(1164,335){\circle{24}}
\put(1176,325){\circle{24}}
\put(1188,323){\circle{24}}
\put(1199,319){\circle{24}}
\put(1210,318){\circle{24}}
\put(1220,312){\circle{24}}
\put(1230,309){\circle{24}}
\put(220,710){\usebox{\plotpoint}}
\multiput(220.00,708.93)(1.267,-0.477){7}{\rule{1.060pt}{0.115pt}}
\multiput(220.00,709.17)(9.800,-5.000){2}{\rule{0.530pt}{0.400pt}}
\multiput(232.00,703.93)(1.378,-0.477){7}{\rule{1.140pt}{0.115pt}}
\multiput(232.00,704.17)(10.634,-5.000){2}{\rule{0.570pt}{0.400pt}}
\multiput(245.00,698.93)(1.267,-0.477){7}{\rule{1.060pt}{0.115pt}}
\multiput(245.00,699.17)(9.800,-5.000){2}{\rule{0.530pt}{0.400pt}}
\multiput(257.00,693.93)(1.267,-0.477){7}{\rule{1.060pt}{0.115pt}}
\multiput(257.00,694.17)(9.800,-5.000){2}{\rule{0.530pt}{0.400pt}}
\multiput(269.00,688.94)(1.651,-0.468){5}{\rule{1.300pt}{0.113pt}}
\multiput(269.00,689.17)(9.302,-4.000){2}{\rule{0.650pt}{0.400pt}}
\multiput(281.00,684.93)(1.378,-0.477){7}{\rule{1.140pt}{0.115pt}}
\multiput(281.00,685.17)(10.634,-5.000){2}{\rule{0.570pt}{0.400pt}}
\multiput(294.00,679.93)(1.267,-0.477){7}{\rule{1.060pt}{0.115pt}}
\multiput(294.00,680.17)(9.800,-5.000){2}{\rule{0.530pt}{0.400pt}}
\multiput(306.00,674.93)(1.267,-0.477){7}{\rule{1.060pt}{0.115pt}}
\multiput(306.00,675.17)(9.800,-5.000){2}{\rule{0.530pt}{0.400pt}}
\multiput(318.00,669.93)(1.378,-0.477){7}{\rule{1.140pt}{0.115pt}}
\multiput(318.00,670.17)(10.634,-5.000){2}{\rule{0.570pt}{0.400pt}}
\multiput(331.00,664.93)(1.267,-0.477){7}{\rule{1.060pt}{0.115pt}}
\multiput(331.00,665.17)(9.800,-5.000){2}{\rule{0.530pt}{0.400pt}}
\multiput(343.00,659.93)(1.267,-0.477){7}{\rule{1.060pt}{0.115pt}}
\multiput(343.00,660.17)(9.800,-5.000){2}{\rule{0.530pt}{0.400pt}}
\multiput(355.00,654.93)(1.267,-0.477){7}{\rule{1.060pt}{0.115pt}}
\multiput(355.00,655.17)(9.800,-5.000){2}{\rule{0.530pt}{0.400pt}}
\multiput(367.00,649.93)(1.378,-0.477){7}{\rule{1.140pt}{0.115pt}}
\multiput(367.00,650.17)(10.634,-5.000){2}{\rule{0.570pt}{0.400pt}}
\multiput(380.00,644.93)(1.267,-0.477){7}{\rule{1.060pt}{0.115pt}}
\multiput(380.00,645.17)(9.800,-5.000){2}{\rule{0.530pt}{0.400pt}}
\multiput(392.00,639.94)(1.651,-0.468){5}{\rule{1.300pt}{0.113pt}}
\multiput(392.00,640.17)(9.302,-4.000){2}{\rule{0.650pt}{0.400pt}}
\multiput(404.00,635.93)(1.378,-0.477){7}{\rule{1.140pt}{0.115pt}}
\multiput(404.00,636.17)(10.634,-5.000){2}{\rule{0.570pt}{0.400pt}}
\multiput(417.00,630.93)(1.267,-0.477){7}{\rule{1.060pt}{0.115pt}}
\multiput(417.00,631.17)(9.800,-5.000){2}{\rule{0.530pt}{0.400pt}}
\multiput(429.00,625.93)(1.267,-0.477){7}{\rule{1.060pt}{0.115pt}}
\multiput(429.00,626.17)(9.800,-5.000){2}{\rule{0.530pt}{0.400pt}}
\multiput(441.00,620.93)(1.267,-0.477){7}{\rule{1.060pt}{0.115pt}}
\multiput(441.00,621.17)(9.800,-5.000){2}{\rule{0.530pt}{0.400pt}}
\multiput(453.00,615.93)(1.378,-0.477){7}{\rule{1.140pt}{0.115pt}}
\multiput(453.00,616.17)(10.634,-5.000){2}{\rule{0.570pt}{0.400pt}}
\multiput(466.00,610.93)(1.267,-0.477){7}{\rule{1.060pt}{0.115pt}}
\multiput(466.00,611.17)(9.800,-5.000){2}{\rule{0.530pt}{0.400pt}}
\multiput(478.00,605.93)(1.267,-0.477){7}{\rule{1.060pt}{0.115pt}}
\multiput(478.00,606.17)(9.800,-5.000){2}{\rule{0.530pt}{0.400pt}}
\multiput(490.00,600.93)(1.378,-0.477){7}{\rule{1.140pt}{0.115pt}}
\multiput(490.00,601.17)(10.634,-5.000){2}{\rule{0.570pt}{0.400pt}}
\multiput(503.00,595.93)(1.267,-0.477){7}{\rule{1.060pt}{0.115pt}}
\multiput(503.00,596.17)(9.800,-5.000){2}{\rule{0.530pt}{0.400pt}}
\multiput(515.00,590.93)(1.267,-0.477){7}{\rule{1.060pt}{0.115pt}}
\multiput(515.00,591.17)(9.800,-5.000){2}{\rule{0.530pt}{0.400pt}}
\multiput(527.00,585.94)(1.651,-0.468){5}{\rule{1.300pt}{0.113pt}}
\multiput(527.00,586.17)(9.302,-4.000){2}{\rule{0.650pt}{0.400pt}}
\multiput(539.00,581.93)(1.378,-0.477){7}{\rule{1.140pt}{0.115pt}}
\multiput(539.00,582.17)(10.634,-5.000){2}{\rule{0.570pt}{0.400pt}}
\multiput(552.00,576.93)(1.267,-0.477){7}{\rule{1.060pt}{0.115pt}}
\multiput(552.00,577.17)(9.800,-5.000){2}{\rule{0.530pt}{0.400pt}}
\multiput(564.00,571.93)(1.267,-0.477){7}{\rule{1.060pt}{0.115pt}}
\multiput(564.00,572.17)(9.800,-5.000){2}{\rule{0.530pt}{0.400pt}}
\multiput(576.00,566.93)(1.267,-0.477){7}{\rule{1.060pt}{0.115pt}}
\multiput(576.00,567.17)(9.800,-5.000){2}{\rule{0.530pt}{0.400pt}}
\multiput(588.00,561.93)(1.378,-0.477){7}{\rule{1.140pt}{0.115pt}}
\multiput(588.00,562.17)(10.634,-5.000){2}{\rule{0.570pt}{0.400pt}}
\multiput(601.00,556.93)(1.267,-0.477){7}{\rule{1.060pt}{0.115pt}}
\multiput(601.00,557.17)(9.800,-5.000){2}{\rule{0.530pt}{0.400pt}}
\multiput(613.00,551.93)(1.267,-0.477){7}{\rule{1.060pt}{0.115pt}}
\multiput(613.00,552.17)(9.800,-5.000){2}{\rule{0.530pt}{0.400pt}}
\multiput(625.00,546.93)(1.378,-0.477){7}{\rule{1.140pt}{0.115pt}}
\multiput(625.00,547.17)(10.634,-5.000){2}{\rule{0.570pt}{0.400pt}}
\multiput(638.00,541.93)(1.267,-0.477){7}{\rule{1.060pt}{0.115pt}}
\multiput(638.00,542.17)(9.800,-5.000){2}{\rule{0.530pt}{0.400pt}}
\multiput(650.00,536.93)(1.267,-0.477){7}{\rule{1.060pt}{0.115pt}}
\multiput(650.00,537.17)(9.800,-5.000){2}{\rule{0.530pt}{0.400pt}}
\multiput(662.00,531.94)(1.651,-0.468){5}{\rule{1.300pt}{0.113pt}}
\multiput(662.00,532.17)(9.302,-4.000){2}{\rule{0.650pt}{0.400pt}}
\multiput(674.00,527.93)(1.378,-0.477){7}{\rule{1.140pt}{0.115pt}}
\multiput(674.00,528.17)(10.634,-5.000){2}{\rule{0.570pt}{0.400pt}}
\multiput(687.00,522.93)(1.267,-0.477){7}{\rule{1.060pt}{0.115pt}}
\multiput(687.00,523.17)(9.800,-5.000){2}{\rule{0.530pt}{0.400pt}}
\multiput(699.00,517.93)(1.267,-0.477){7}{\rule{1.060pt}{0.115pt}}
\multiput(699.00,518.17)(9.800,-5.000){2}{\rule{0.530pt}{0.400pt}}
\multiput(711.00,512.93)(1.378,-0.477){7}{\rule{1.140pt}{0.115pt}}
\multiput(711.00,513.17)(10.634,-5.000){2}{\rule{0.570pt}{0.400pt}}
\multiput(724.00,507.93)(1.267,-0.477){7}{\rule{1.060pt}{0.115pt}}
\multiput(724.00,508.17)(9.800,-5.000){2}{\rule{0.530pt}{0.400pt}}
\multiput(736.00,502.93)(1.267,-0.477){7}{\rule{1.060pt}{0.115pt}}
\multiput(736.00,503.17)(9.800,-5.000){2}{\rule{0.530pt}{0.400pt}}
\multiput(748.00,497.93)(1.267,-0.477){7}{\rule{1.060pt}{0.115pt}}
\multiput(748.00,498.17)(9.800,-5.000){2}{\rule{0.530pt}{0.400pt}}
\multiput(760.00,492.93)(1.378,-0.477){7}{\rule{1.140pt}{0.115pt}}
\multiput(760.00,493.17)(10.634,-5.000){2}{\rule{0.570pt}{0.400pt}}
\multiput(773.00,487.93)(1.267,-0.477){7}{\rule{1.060pt}{0.115pt}}
\multiput(773.00,488.17)(9.800,-5.000){2}{\rule{0.530pt}{0.400pt}}
\multiput(785.00,482.93)(1.267,-0.477){7}{\rule{1.060pt}{0.115pt}}
\multiput(785.00,483.17)(9.800,-5.000){2}{\rule{0.530pt}{0.400pt}}
\multiput(797.00,477.94)(1.797,-0.468){5}{\rule{1.400pt}{0.113pt}}
\multiput(797.00,478.17)(10.094,-4.000){2}{\rule{0.700pt}{0.400pt}}
\multiput(810.00,473.93)(1.267,-0.477){7}{\rule{1.060pt}{0.115pt}}
\multiput(810.00,474.17)(9.800,-5.000){2}{\rule{0.530pt}{0.400pt}}
\multiput(822.00,468.93)(1.267,-0.477){7}{\rule{1.060pt}{0.115pt}}
\multiput(822.00,469.17)(9.800,-5.000){2}{\rule{0.530pt}{0.400pt}}
\multiput(834.00,463.93)(1.267,-0.477){7}{\rule{1.060pt}{0.115pt}}
\multiput(834.00,464.17)(9.800,-5.000){2}{\rule{0.530pt}{0.400pt}}
\multiput(846.00,458.93)(1.378,-0.477){7}{\rule{1.140pt}{0.115pt}}
\multiput(846.00,459.17)(10.634,-5.000){2}{\rule{0.570pt}{0.400pt}}
\multiput(859.00,453.93)(1.267,-0.477){7}{\rule{1.060pt}{0.115pt}}
\multiput(859.00,454.17)(9.800,-5.000){2}{\rule{0.530pt}{0.400pt}}
\multiput(871.00,448.93)(1.267,-0.477){7}{\rule{1.060pt}{0.115pt}}
\multiput(871.00,449.17)(9.800,-5.000){2}{\rule{0.530pt}{0.400pt}}
\multiput(883.00,443.93)(1.378,-0.477){7}{\rule{1.140pt}{0.115pt}}
\multiput(883.00,444.17)(10.634,-5.000){2}{\rule{0.570pt}{0.400pt}}
\multiput(896.00,438.93)(1.267,-0.477){7}{\rule{1.060pt}{0.115pt}}
\multiput(896.00,439.17)(9.800,-5.000){2}{\rule{0.530pt}{0.400pt}}
\multiput(908.00,433.93)(1.267,-0.477){7}{\rule{1.060pt}{0.115pt}}
\multiput(908.00,434.17)(9.800,-5.000){2}{\rule{0.530pt}{0.400pt}}
\multiput(920.00,428.93)(1.267,-0.477){7}{\rule{1.060pt}{0.115pt}}
\multiput(920.00,429.17)(9.800,-5.000){2}{\rule{0.530pt}{0.400pt}}
\multiput(932.00,423.94)(1.797,-0.468){5}{\rule{1.400pt}{0.113pt}}
\multiput(932.00,424.17)(10.094,-4.000){2}{\rule{0.700pt}{0.400pt}}
\multiput(945.00,419.93)(1.267,-0.477){7}{\rule{1.060pt}{0.115pt}}
\multiput(945.00,420.17)(9.800,-5.000){2}{\rule{0.530pt}{0.400pt}}
\multiput(957.00,414.93)(1.267,-0.477){7}{\rule{1.060pt}{0.115pt}}
\multiput(957.00,415.17)(9.800,-5.000){2}{\rule{0.530pt}{0.400pt}}
\multiput(969.00,409.93)(1.378,-0.477){7}{\rule{1.140pt}{0.115pt}}
\multiput(969.00,410.17)(10.634,-5.000){2}{\rule{0.570pt}{0.400pt}}
\multiput(982.00,404.93)(1.267,-0.477){7}{\rule{1.060pt}{0.115pt}}
\multiput(982.00,405.17)(9.800,-5.000){2}{\rule{0.530pt}{0.400pt}}
\multiput(994.00,399.93)(1.267,-0.477){7}{\rule{1.060pt}{0.115pt}}
\multiput(994.00,400.17)(9.800,-5.000){2}{\rule{0.530pt}{0.400pt}}
\multiput(1006.00,394.93)(1.267,-0.477){7}{\rule{1.060pt}{0.115pt}}
\multiput(1006.00,395.17)(9.800,-5.000){2}{\rule{0.530pt}{0.400pt}}
\multiput(1018.00,389.93)(1.378,-0.477){7}{\rule{1.140pt}{0.115pt}}
\multiput(1018.00,390.17)(10.634,-5.000){2}{\rule{0.570pt}{0.400pt}}
\multiput(1031.00,384.93)(1.267,-0.477){7}{\rule{1.060pt}{0.115pt}}
\multiput(1031.00,385.17)(9.800,-5.000){2}{\rule{0.530pt}{0.400pt}}
\multiput(1043.00,379.93)(1.267,-0.477){7}{\rule{1.060pt}{0.115pt}}
\multiput(1043.00,380.17)(9.800,-5.000){2}{\rule{0.530pt}{0.400pt}}
\multiput(1055.00,374.93)(1.378,-0.477){7}{\rule{1.140pt}{0.115pt}}
\multiput(1055.00,375.17)(10.634,-5.000){2}{\rule{0.570pt}{0.400pt}}
\multiput(1068.00,369.94)(1.651,-0.468){5}{\rule{1.300pt}{0.113pt}}
\multiput(1068.00,370.17)(9.302,-4.000){2}{\rule{0.650pt}{0.400pt}}
\multiput(1080.00,365.93)(1.267,-0.477){7}{\rule{1.060pt}{0.115pt}}
\multiput(1080.00,366.17)(9.800,-5.000){2}{\rule{0.530pt}{0.400pt}}
\multiput(1092.00,360.93)(1.267,-0.477){7}{\rule{1.060pt}{0.115pt}}
\multiput(1092.00,361.17)(9.800,-5.000){2}{\rule{0.530pt}{0.400pt}}
\multiput(1104.00,355.93)(1.378,-0.477){7}{\rule{1.140pt}{0.115pt}}
\multiput(1104.00,356.17)(10.634,-5.000){2}{\rule{0.570pt}{0.400pt}}
\multiput(1117.00,350.93)(1.267,-0.477){7}{\rule{1.060pt}{0.115pt}}
\multiput(1117.00,351.17)(9.800,-5.000){2}{\rule{0.530pt}{0.400pt}}
\multiput(1129.00,345.93)(1.267,-0.477){7}{\rule{1.060pt}{0.115pt}}
\multiput(1129.00,346.17)(9.800,-5.000){2}{\rule{0.530pt}{0.400pt}}
\multiput(1141.00,340.93)(1.267,-0.477){7}{\rule{1.060pt}{0.115pt}}
\multiput(1141.00,341.17)(9.800,-5.000){2}{\rule{0.530pt}{0.400pt}}
\multiput(1153.00,335.93)(1.378,-0.477){7}{\rule{1.140pt}{0.115pt}}
\multiput(1153.00,336.17)(10.634,-5.000){2}{\rule{0.570pt}{0.400pt}}
\multiput(1166.00,330.93)(1.267,-0.477){7}{\rule{1.060pt}{0.115pt}}
\multiput(1166.00,331.17)(9.800,-5.000){2}{\rule{0.530pt}{0.400pt}}
\multiput(1178.00,325.93)(1.267,-0.477){7}{\rule{1.060pt}{0.115pt}}
\multiput(1178.00,326.17)(9.800,-5.000){2}{\rule{0.530pt}{0.400pt}}
\multiput(1190.00,320.93)(1.378,-0.477){7}{\rule{1.140pt}{0.115pt}}
\multiput(1190.00,321.17)(10.634,-5.000){2}{\rule{0.570pt}{0.400pt}}
\multiput(1203.00,315.94)(1.651,-0.468){5}{\rule{1.300pt}{0.113pt}}
\multiput(1203.00,316.17)(9.302,-4.000){2}{\rule{0.650pt}{0.400pt}}
\multiput(1215.00,311.93)(1.267,-0.477){7}{\rule{1.060pt}{0.115pt}}
\multiput(1215.00,312.17)(9.800,-5.000){2}{\rule{0.530pt}{0.400pt}}
\multiput(1227.00,306.93)(1.267,-0.477){7}{\rule{1.060pt}{0.115pt}}
\multiput(1227.00,307.17)(9.800,-5.000){2}{\rule{0.530pt}{0.400pt}}
\multiput(1239.00,301.93)(1.378,-0.477){7}{\rule{1.140pt}{0.115pt}}
\multiput(1239.00,302.17)(10.634,-5.000){2}{\rule{0.570pt}{0.400pt}}
\multiput(1252.00,296.93)(1.267,-0.477){7}{\rule{1.060pt}{0.115pt}}
\multiput(1252.00,297.17)(9.800,-5.000){2}{\rule{0.530pt}{0.400pt}}
\multiput(1264.00,291.93)(1.267,-0.477){7}{\rule{1.060pt}{0.115pt}}
\multiput(1264.00,292.17)(9.800,-5.000){2}{\rule{0.530pt}{0.400pt}}
\multiput(1276.00,286.93)(1.378,-0.477){7}{\rule{1.140pt}{0.115pt}}
\multiput(1276.00,287.17)(10.634,-5.000){2}{\rule{0.570pt}{0.400pt}}
\multiput(1289.00,281.93)(1.267,-0.477){7}{\rule{1.060pt}{0.115pt}}
\multiput(1289.00,282.17)(9.800,-5.000){2}{\rule{0.530pt}{0.400pt}}
\multiput(1301.00,276.93)(1.267,-0.477){7}{\rule{1.060pt}{0.115pt}}
\multiput(1301.00,277.17)(9.800,-5.000){2}{\rule{0.530pt}{0.400pt}}
\multiput(1313.00,271.93)(1.267,-0.477){7}{\rule{1.060pt}{0.115pt}}
\multiput(1313.00,272.17)(9.800,-5.000){2}{\rule{0.530pt}{0.400pt}}
\multiput(1325.00,266.93)(1.378,-0.477){7}{\rule{1.140pt}{0.115pt}}
\multiput(1325.00,267.17)(10.634,-5.000){2}{\rule{0.570pt}{0.400pt}}
\multiput(1338.00,261.94)(1.651,-0.468){5}{\rule{1.300pt}{0.113pt}}
\multiput(1338.00,262.17)(9.302,-4.000){2}{\rule{0.650pt}{0.400pt}}
\multiput(1350.00,257.93)(1.267,-0.477){7}{\rule{1.060pt}{0.115pt}}
\multiput(1350.00,258.17)(9.800,-5.000){2}{\rule{0.530pt}{0.400pt}}
\multiput(1362.00,252.93)(1.378,-0.477){7}{\rule{1.140pt}{0.115pt}}
\multiput(1362.00,253.17)(10.634,-5.000){2}{\rule{0.570pt}{0.400pt}}
\multiput(1375.00,247.93)(1.267,-0.477){7}{\rule{1.060pt}{0.115pt}}
\multiput(1375.00,248.17)(9.800,-5.000){2}{\rule{0.530pt}{0.400pt}}
\multiput(1387.00,242.93)(1.267,-0.477){7}{\rule{1.060pt}{0.115pt}}
\multiput(1387.00,243.17)(9.800,-5.000){2}{\rule{0.530pt}{0.400pt}}
\multiput(1399.00,237.93)(1.267,-0.477){7}{\rule{1.060pt}{0.115pt}}
\multiput(1399.00,238.17)(9.800,-5.000){2}{\rule{0.530pt}{0.400pt}}
\multiput(1411.00,232.93)(1.378,-0.477){7}{\rule{1.140pt}{0.115pt}}
\multiput(1411.00,233.17)(10.634,-5.000){2}{\rule{0.570pt}{0.400pt}}
\multiput(1424.00,227.93)(1.267,-0.477){7}{\rule{1.060pt}{0.115pt}}
\multiput(1424.00,228.17)(9.800,-5.000){2}{\rule{0.530pt}{0.400pt}}
\end{picture}
\end{center}
\end{figure}

\clearpage
\begin{figure}
\begin{center}
% GNUPLOT: LaTeX picture
\setlength{\unitlength}{0.240900pt}
\ifx\plotpoint\undefined\newsavebox{\plotpoint}\fi
\sbox{\plotpoint}{\rule[-0.200pt]{0.400pt}{0.400pt}}%
\begin{picture}(1500,900)(0,0)
\font\gnuplot=cmr10 at 10pt
\gnuplot
\sbox{\plotpoint}{\rule[-0.200pt]{0.400pt}{0.400pt}}%
\put(220.0,266.0){\rule[-0.200pt]{4.818pt}{0.400pt}}
\put(198,266){\makebox(0,0)[r]{{\large $1$}}}
\put(1416.0,266.0){\rule[-0.200pt]{4.818pt}{0.400pt}}
\put(220.0,419.0){\rule[-0.200pt]{4.818pt}{0.400pt}}
\put(198,419){\makebox(0,0)[r]{{\large $10$}}}
\put(1416.0,419.0){\rule[-0.200pt]{4.818pt}{0.400pt}}
\put(220.0,571.0){\rule[-0.200pt]{4.818pt}{0.400pt}}
\put(198,571){\makebox(0,0)[r]{{\large $100$}}}
\put(1416.0,571.0){\rule[-0.200pt]{4.818pt}{0.400pt}}
\put(220.0,724.0){\rule[-0.200pt]{4.818pt}{0.400pt}}
\put(198,724){\makebox(0,0)[r]{{\large $1000$}}}
\put(1416.0,724.0){\rule[-0.200pt]{4.818pt}{0.400pt}}
\put(426.0,113.0){\rule[-0.200pt]{0.400pt}{4.818pt}}
\put(426,68){\makebox(0,0){{\large $4$}}}
\put(426.0,857.0){\rule[-0.200pt]{0.400pt}{4.818pt}}
\put(632.0,113.0){\rule[-0.200pt]{0.400pt}{4.818pt}}
\put(632,68){\makebox(0,0){{\large $8$}}}
\put(632.0,857.0){\rule[-0.200pt]{0.400pt}{4.818pt}}
\put(838.0,113.0){\rule[-0.200pt]{0.400pt}{4.818pt}}
\put(838,68){\makebox(0,0){{\large $16$}}}
\put(838.0,857.0){\rule[-0.200pt]{0.400pt}{4.818pt}}
\put(1043.0,113.0){\rule[-0.200pt]{0.400pt}{4.818pt}}
\put(1043,68){\makebox(0,0){{\large $32$}}}
\put(1043.0,857.0){\rule[-0.200pt]{0.400pt}{4.818pt}}
\put(1230.0,113.0){\rule[-0.200pt]{0.400pt}{4.818pt}}
\put(1230,68){\makebox(0,0){{\large $60$}}}
\put(1230.0,857.0){\rule[-0.200pt]{0.400pt}{4.818pt}}
\put(220.0,113.0){\rule[-0.200pt]{292.934pt}{0.400pt}}
\put(1436.0,113.0){\rule[-0.200pt]{0.400pt}{184.048pt}}
\put(220.0,877.0){\rule[-0.200pt]{292.934pt}{0.400pt}}
\put(45,495){\makebox(0,0){{\Large $\chi^{\dagger \prime}$}}}
\put(828,-22){\makebox(0,0){{\Large $L$}}}
\put(818,-344){\makebox(0,0){{\large Figure 2}}}
\put(818,-497){\makebox(0,0){{\large Atsushi Yamagata}}}
\put(220.0,113.0){\rule[-0.200pt]{0.400pt}{184.048pt}}
\put(426,335){\circle{24}}
\put(546,392){\circle{24}}
\put(632,431){\circle{24}}
\put(698,462){\circle{24}}
\put(752,487){\circle{24}}
\put(798,508){\circle{24}}
\put(838,526){\circle{24}}
\put(873,542){\circle{24}}
\put(904,557){\circle{24}}
\put(932,571){\circle{24}}
\put(958,583){\circle{24}}
\put(982,594){\circle{24}}
\put(1004,605){\circle{24}}
\put(1024,614){\circle{24}}
\put(1043,624){\circle{24}}
\put(1061,634){\circle{24}}
\put(1078,641){\circle{24}}
\put(1094,649){\circle{24}}
\put(1110,657){\circle{24}}
\put(1124,666){\circle{24}}
\put(1138,674){\circle{24}}
\put(1151,680){\circle{24}}
\put(1164,687){\circle{24}}
\put(1176,692){\circle{24}}
\put(1188,699){\circle{24}}
\put(1199,705){\circle{24}}
\put(1210,712){\circle{24}}
\put(1220,716){\circle{24}}
\put(1230,722){\circle{24}}
\put(220,186){\usebox{\plotpoint}}
\multiput(220.00,186.59)(1.033,0.482){9}{\rule{0.900pt}{0.116pt}}
\multiput(220.00,185.17)(10.132,6.000){2}{\rule{0.450pt}{0.400pt}}
\multiput(232.00,192.59)(0.950,0.485){11}{\rule{0.843pt}{0.117pt}}
\multiput(232.00,191.17)(11.251,7.000){2}{\rule{0.421pt}{0.400pt}}
\multiput(245.00,199.59)(1.033,0.482){9}{\rule{0.900pt}{0.116pt}}
\multiput(245.00,198.17)(10.132,6.000){2}{\rule{0.450pt}{0.400pt}}
\multiput(257.00,205.59)(0.874,0.485){11}{\rule{0.786pt}{0.117pt}}
\multiput(257.00,204.17)(10.369,7.000){2}{\rule{0.393pt}{0.400pt}}
\multiput(269.00,212.59)(1.033,0.482){9}{\rule{0.900pt}{0.116pt}}
\multiput(269.00,211.17)(10.132,6.000){2}{\rule{0.450pt}{0.400pt}}
\multiput(281.00,218.59)(0.950,0.485){11}{\rule{0.843pt}{0.117pt}}
\multiput(281.00,217.17)(11.251,7.000){2}{\rule{0.421pt}{0.400pt}}
\multiput(294.00,225.59)(1.033,0.482){9}{\rule{0.900pt}{0.116pt}}
\multiput(294.00,224.17)(10.132,6.000){2}{\rule{0.450pt}{0.400pt}}
\multiput(306.00,231.59)(0.874,0.485){11}{\rule{0.786pt}{0.117pt}}
\multiput(306.00,230.17)(10.369,7.000){2}{\rule{0.393pt}{0.400pt}}
\multiput(318.00,238.59)(1.123,0.482){9}{\rule{0.967pt}{0.116pt}}
\multiput(318.00,237.17)(10.994,6.000){2}{\rule{0.483pt}{0.400pt}}
\multiput(331.00,244.59)(0.874,0.485){11}{\rule{0.786pt}{0.117pt}}
\multiput(331.00,243.17)(10.369,7.000){2}{\rule{0.393pt}{0.400pt}}
\multiput(343.00,251.59)(1.033,0.482){9}{\rule{0.900pt}{0.116pt}}
\multiput(343.00,250.17)(10.132,6.000){2}{\rule{0.450pt}{0.400pt}}
\multiput(355.00,257.59)(0.874,0.485){11}{\rule{0.786pt}{0.117pt}}
\multiput(355.00,256.17)(10.369,7.000){2}{\rule{0.393pt}{0.400pt}}
\multiput(367.00,264.59)(1.123,0.482){9}{\rule{0.967pt}{0.116pt}}
\multiput(367.00,263.17)(10.994,6.000){2}{\rule{0.483pt}{0.400pt}}
\multiput(380.00,270.59)(0.874,0.485){11}{\rule{0.786pt}{0.117pt}}
\multiput(380.00,269.17)(10.369,7.000){2}{\rule{0.393pt}{0.400pt}}
\multiput(392.00,277.59)(1.033,0.482){9}{\rule{0.900pt}{0.116pt}}
\multiput(392.00,276.17)(10.132,6.000){2}{\rule{0.450pt}{0.400pt}}
\multiput(404.00,283.59)(0.950,0.485){11}{\rule{0.843pt}{0.117pt}}
\multiput(404.00,282.17)(11.251,7.000){2}{\rule{0.421pt}{0.400pt}}
\multiput(417.00,290.59)(1.033,0.482){9}{\rule{0.900pt}{0.116pt}}
\multiput(417.00,289.17)(10.132,6.000){2}{\rule{0.450pt}{0.400pt}}
\multiput(429.00,296.59)(0.874,0.485){11}{\rule{0.786pt}{0.117pt}}
\multiput(429.00,295.17)(10.369,7.000){2}{\rule{0.393pt}{0.400pt}}
\multiput(441.00,303.59)(1.033,0.482){9}{\rule{0.900pt}{0.116pt}}
\multiput(441.00,302.17)(10.132,6.000){2}{\rule{0.450pt}{0.400pt}}
\multiput(453.00,309.59)(0.950,0.485){11}{\rule{0.843pt}{0.117pt}}
\multiput(453.00,308.17)(11.251,7.000){2}{\rule{0.421pt}{0.400pt}}
\multiput(466.00,316.59)(1.033,0.482){9}{\rule{0.900pt}{0.116pt}}
\multiput(466.00,315.17)(10.132,6.000){2}{\rule{0.450pt}{0.400pt}}
\multiput(478.00,322.59)(0.874,0.485){11}{\rule{0.786pt}{0.117pt}}
\multiput(478.00,321.17)(10.369,7.000){2}{\rule{0.393pt}{0.400pt}}
\multiput(490.00,329.59)(1.123,0.482){9}{\rule{0.967pt}{0.116pt}}
\multiput(490.00,328.17)(10.994,6.000){2}{\rule{0.483pt}{0.400pt}}
\multiput(503.00,335.59)(0.874,0.485){11}{\rule{0.786pt}{0.117pt}}
\multiput(503.00,334.17)(10.369,7.000){2}{\rule{0.393pt}{0.400pt}}
\multiput(515.00,342.59)(1.033,0.482){9}{\rule{0.900pt}{0.116pt}}
\multiput(515.00,341.17)(10.132,6.000){2}{\rule{0.450pt}{0.400pt}}
\multiput(527.00,348.59)(0.874,0.485){11}{\rule{0.786pt}{0.117pt}}
\multiput(527.00,347.17)(10.369,7.000){2}{\rule{0.393pt}{0.400pt}}
\multiput(539.00,355.59)(1.123,0.482){9}{\rule{0.967pt}{0.116pt}}
\multiput(539.00,354.17)(10.994,6.000){2}{\rule{0.483pt}{0.400pt}}
\multiput(552.00,361.59)(0.874,0.485){11}{\rule{0.786pt}{0.117pt}}
\multiput(552.00,360.17)(10.369,7.000){2}{\rule{0.393pt}{0.400pt}}
\multiput(564.00,368.59)(1.033,0.482){9}{\rule{0.900pt}{0.116pt}}
\multiput(564.00,367.17)(10.132,6.000){2}{\rule{0.450pt}{0.400pt}}
\multiput(576.00,374.59)(0.874,0.485){11}{\rule{0.786pt}{0.117pt}}
\multiput(576.00,373.17)(10.369,7.000){2}{\rule{0.393pt}{0.400pt}}
\multiput(588.00,381.59)(0.950,0.485){11}{\rule{0.843pt}{0.117pt}}
\multiput(588.00,380.17)(11.251,7.000){2}{\rule{0.421pt}{0.400pt}}
\multiput(601.00,388.59)(1.033,0.482){9}{\rule{0.900pt}{0.116pt}}
\multiput(601.00,387.17)(10.132,6.000){2}{\rule{0.450pt}{0.400pt}}
\multiput(613.00,394.59)(0.874,0.485){11}{\rule{0.786pt}{0.117pt}}
\multiput(613.00,393.17)(10.369,7.000){2}{\rule{0.393pt}{0.400pt}}
\multiput(625.00,401.59)(1.123,0.482){9}{\rule{0.967pt}{0.116pt}}
\multiput(625.00,400.17)(10.994,6.000){2}{\rule{0.483pt}{0.400pt}}
\multiput(638.00,407.59)(0.874,0.485){11}{\rule{0.786pt}{0.117pt}}
\multiput(638.00,406.17)(10.369,7.000){2}{\rule{0.393pt}{0.400pt}}
\multiput(650.00,414.59)(1.033,0.482){9}{\rule{0.900pt}{0.116pt}}
\multiput(650.00,413.17)(10.132,6.000){2}{\rule{0.450pt}{0.400pt}}
\multiput(662.00,420.59)(0.874,0.485){11}{\rule{0.786pt}{0.117pt}}
\multiput(662.00,419.17)(10.369,7.000){2}{\rule{0.393pt}{0.400pt}}
\multiput(674.00,427.59)(1.123,0.482){9}{\rule{0.967pt}{0.116pt}}
\multiput(674.00,426.17)(10.994,6.000){2}{\rule{0.483pt}{0.400pt}}
\multiput(687.00,433.59)(0.874,0.485){11}{\rule{0.786pt}{0.117pt}}
\multiput(687.00,432.17)(10.369,7.000){2}{\rule{0.393pt}{0.400pt}}
\multiput(699.00,440.59)(1.033,0.482){9}{\rule{0.900pt}{0.116pt}}
\multiput(699.00,439.17)(10.132,6.000){2}{\rule{0.450pt}{0.400pt}}
\multiput(711.00,446.59)(0.950,0.485){11}{\rule{0.843pt}{0.117pt}}
\multiput(711.00,445.17)(11.251,7.000){2}{\rule{0.421pt}{0.400pt}}
\multiput(724.00,453.59)(1.033,0.482){9}{\rule{0.900pt}{0.116pt}}
\multiput(724.00,452.17)(10.132,6.000){2}{\rule{0.450pt}{0.400pt}}
\multiput(736.00,459.59)(0.874,0.485){11}{\rule{0.786pt}{0.117pt}}
\multiput(736.00,458.17)(10.369,7.000){2}{\rule{0.393pt}{0.400pt}}
\multiput(748.00,466.59)(1.033,0.482){9}{\rule{0.900pt}{0.116pt}}
\multiput(748.00,465.17)(10.132,6.000){2}{\rule{0.450pt}{0.400pt}}
\multiput(760.00,472.59)(0.950,0.485){11}{\rule{0.843pt}{0.117pt}}
\multiput(760.00,471.17)(11.251,7.000){2}{\rule{0.421pt}{0.400pt}}
\multiput(773.00,479.59)(1.033,0.482){9}{\rule{0.900pt}{0.116pt}}
\multiput(773.00,478.17)(10.132,6.000){2}{\rule{0.450pt}{0.400pt}}
\multiput(785.00,485.59)(0.874,0.485){11}{\rule{0.786pt}{0.117pt}}
\multiput(785.00,484.17)(10.369,7.000){2}{\rule{0.393pt}{0.400pt}}
\multiput(797.00,492.59)(1.123,0.482){9}{\rule{0.967pt}{0.116pt}}
\multiput(797.00,491.17)(10.994,6.000){2}{\rule{0.483pt}{0.400pt}}
\multiput(810.00,498.59)(0.874,0.485){11}{\rule{0.786pt}{0.117pt}}
\multiput(810.00,497.17)(10.369,7.000){2}{\rule{0.393pt}{0.400pt}}
\multiput(822.00,505.59)(1.033,0.482){9}{\rule{0.900pt}{0.116pt}}
\multiput(822.00,504.17)(10.132,6.000){2}{\rule{0.450pt}{0.400pt}}
\multiput(834.00,511.59)(0.874,0.485){11}{\rule{0.786pt}{0.117pt}}
\multiput(834.00,510.17)(10.369,7.000){2}{\rule{0.393pt}{0.400pt}}
\multiput(846.00,518.59)(1.123,0.482){9}{\rule{0.967pt}{0.116pt}}
\multiput(846.00,517.17)(10.994,6.000){2}{\rule{0.483pt}{0.400pt}}
\multiput(859.00,524.59)(0.874,0.485){11}{\rule{0.786pt}{0.117pt}}
\multiput(859.00,523.17)(10.369,7.000){2}{\rule{0.393pt}{0.400pt}}
\multiput(871.00,531.59)(1.033,0.482){9}{\rule{0.900pt}{0.116pt}}
\multiput(871.00,530.17)(10.132,6.000){2}{\rule{0.450pt}{0.400pt}}
\multiput(883.00,537.59)(0.950,0.485){11}{\rule{0.843pt}{0.117pt}}
\multiput(883.00,536.17)(11.251,7.000){2}{\rule{0.421pt}{0.400pt}}
\multiput(896.00,544.59)(1.033,0.482){9}{\rule{0.900pt}{0.116pt}}
\multiput(896.00,543.17)(10.132,6.000){2}{\rule{0.450pt}{0.400pt}}
\multiput(908.00,550.59)(0.874,0.485){11}{\rule{0.786pt}{0.117pt}}
\multiput(908.00,549.17)(10.369,7.000){2}{\rule{0.393pt}{0.400pt}}
\multiput(920.00,557.59)(1.033,0.482){9}{\rule{0.900pt}{0.116pt}}
\multiput(920.00,556.17)(10.132,6.000){2}{\rule{0.450pt}{0.400pt}}
\multiput(932.00,563.59)(0.950,0.485){11}{\rule{0.843pt}{0.117pt}}
\multiput(932.00,562.17)(11.251,7.000){2}{\rule{0.421pt}{0.400pt}}
\multiput(945.00,570.59)(1.033,0.482){9}{\rule{0.900pt}{0.116pt}}
\multiput(945.00,569.17)(10.132,6.000){2}{\rule{0.450pt}{0.400pt}}
\multiput(957.00,576.59)(0.874,0.485){11}{\rule{0.786pt}{0.117pt}}
\multiput(957.00,575.17)(10.369,7.000){2}{\rule{0.393pt}{0.400pt}}
\multiput(969.00,583.59)(1.123,0.482){9}{\rule{0.967pt}{0.116pt}}
\multiput(969.00,582.17)(10.994,6.000){2}{\rule{0.483pt}{0.400pt}}
\multiput(982.00,589.59)(0.874,0.485){11}{\rule{0.786pt}{0.117pt}}
\multiput(982.00,588.17)(10.369,7.000){2}{\rule{0.393pt}{0.400pt}}
\multiput(994.00,596.59)(1.033,0.482){9}{\rule{0.900pt}{0.116pt}}
\multiput(994.00,595.17)(10.132,6.000){2}{\rule{0.450pt}{0.400pt}}
\multiput(1006.00,602.59)(0.874,0.485){11}{\rule{0.786pt}{0.117pt}}
\multiput(1006.00,601.17)(10.369,7.000){2}{\rule{0.393pt}{0.400pt}}
\multiput(1018.00,609.59)(1.123,0.482){9}{\rule{0.967pt}{0.116pt}}
\multiput(1018.00,608.17)(10.994,6.000){2}{\rule{0.483pt}{0.400pt}}
\multiput(1031.00,615.59)(0.874,0.485){11}{\rule{0.786pt}{0.117pt}}
\multiput(1031.00,614.17)(10.369,7.000){2}{\rule{0.393pt}{0.400pt}}
\multiput(1043.00,622.59)(1.033,0.482){9}{\rule{0.900pt}{0.116pt}}
\multiput(1043.00,621.17)(10.132,6.000){2}{\rule{0.450pt}{0.400pt}}
\multiput(1055.00,628.59)(0.950,0.485){11}{\rule{0.843pt}{0.117pt}}
\multiput(1055.00,627.17)(11.251,7.000){2}{\rule{0.421pt}{0.400pt}}
\multiput(1068.00,635.59)(1.033,0.482){9}{\rule{0.900pt}{0.116pt}}
\multiput(1068.00,634.17)(10.132,6.000){2}{\rule{0.450pt}{0.400pt}}
\multiput(1080.00,641.59)(0.874,0.485){11}{\rule{0.786pt}{0.117pt}}
\multiput(1080.00,640.17)(10.369,7.000){2}{\rule{0.393pt}{0.400pt}}
\multiput(1092.00,648.59)(1.033,0.482){9}{\rule{0.900pt}{0.116pt}}
\multiput(1092.00,647.17)(10.132,6.000){2}{\rule{0.450pt}{0.400pt}}
\multiput(1104.00,654.59)(0.950,0.485){11}{\rule{0.843pt}{0.117pt}}
\multiput(1104.00,653.17)(11.251,7.000){2}{\rule{0.421pt}{0.400pt}}
\multiput(1117.00,661.59)(1.033,0.482){9}{\rule{0.900pt}{0.116pt}}
\multiput(1117.00,660.17)(10.132,6.000){2}{\rule{0.450pt}{0.400pt}}
\multiput(1129.00,667.59)(0.874,0.485){11}{\rule{0.786pt}{0.117pt}}
\multiput(1129.00,666.17)(10.369,7.000){2}{\rule{0.393pt}{0.400pt}}
\multiput(1141.00,674.59)(1.033,0.482){9}{\rule{0.900pt}{0.116pt}}
\multiput(1141.00,673.17)(10.132,6.000){2}{\rule{0.450pt}{0.400pt}}
\multiput(1153.00,680.59)(0.950,0.485){11}{\rule{0.843pt}{0.117pt}}
\multiput(1153.00,679.17)(11.251,7.000){2}{\rule{0.421pt}{0.400pt}}
\multiput(1166.00,687.59)(1.033,0.482){9}{\rule{0.900pt}{0.116pt}}
\multiput(1166.00,686.17)(10.132,6.000){2}{\rule{0.450pt}{0.400pt}}
\multiput(1178.00,693.59)(0.874,0.485){11}{\rule{0.786pt}{0.117pt}}
\multiput(1178.00,692.17)(10.369,7.000){2}{\rule{0.393pt}{0.400pt}}
\multiput(1190.00,700.59)(1.123,0.482){9}{\rule{0.967pt}{0.116pt}}
\multiput(1190.00,699.17)(10.994,6.000){2}{\rule{0.483pt}{0.400pt}}
\multiput(1203.00,706.59)(0.874,0.485){11}{\rule{0.786pt}{0.117pt}}
\multiput(1203.00,705.17)(10.369,7.000){2}{\rule{0.393pt}{0.400pt}}
\multiput(1215.00,713.59)(1.033,0.482){9}{\rule{0.900pt}{0.116pt}}
\multiput(1215.00,712.17)(10.132,6.000){2}{\rule{0.450pt}{0.400pt}}
\multiput(1227.00,719.59)(0.874,0.485){11}{\rule{0.786pt}{0.117pt}}
\multiput(1227.00,718.17)(10.369,7.000){2}{\rule{0.393pt}{0.400pt}}
\multiput(1239.00,726.59)(1.123,0.482){9}{\rule{0.967pt}{0.116pt}}
\multiput(1239.00,725.17)(10.994,6.000){2}{\rule{0.483pt}{0.400pt}}
\multiput(1252.00,732.59)(0.874,0.485){11}{\rule{0.786pt}{0.117pt}}
\multiput(1252.00,731.17)(10.369,7.000){2}{\rule{0.393pt}{0.400pt}}
\multiput(1264.00,739.59)(1.033,0.482){9}{\rule{0.900pt}{0.116pt}}
\multiput(1264.00,738.17)(10.132,6.000){2}{\rule{0.450pt}{0.400pt}}
\multiput(1276.00,745.59)(0.950,0.485){11}{\rule{0.843pt}{0.117pt}}
\multiput(1276.00,744.17)(11.251,7.000){2}{\rule{0.421pt}{0.400pt}}
\multiput(1289.00,752.59)(1.033,0.482){9}{\rule{0.900pt}{0.116pt}}
\multiput(1289.00,751.17)(10.132,6.000){2}{\rule{0.450pt}{0.400pt}}
\multiput(1301.00,758.59)(0.874,0.485){11}{\rule{0.786pt}{0.117pt}}
\multiput(1301.00,757.17)(10.369,7.000){2}{\rule{0.393pt}{0.400pt}}
\multiput(1313.00,765.59)(1.033,0.482){9}{\rule{0.900pt}{0.116pt}}
\multiput(1313.00,764.17)(10.132,6.000){2}{\rule{0.450pt}{0.400pt}}
\multiput(1325.00,771.59)(0.950,0.485){11}{\rule{0.843pt}{0.117pt}}
\multiput(1325.00,770.17)(11.251,7.000){2}{\rule{0.421pt}{0.400pt}}
\multiput(1338.00,778.59)(1.033,0.482){9}{\rule{0.900pt}{0.116pt}}
\multiput(1338.00,777.17)(10.132,6.000){2}{\rule{0.450pt}{0.400pt}}
\multiput(1350.00,784.59)(0.874,0.485){11}{\rule{0.786pt}{0.117pt}}
\multiput(1350.00,783.17)(10.369,7.000){2}{\rule{0.393pt}{0.400pt}}
\multiput(1362.00,791.59)(1.123,0.482){9}{\rule{0.967pt}{0.116pt}}
\multiput(1362.00,790.17)(10.994,6.000){2}{\rule{0.483pt}{0.400pt}}
\multiput(1375.00,797.59)(0.874,0.485){11}{\rule{0.786pt}{0.117pt}}
\multiput(1375.00,796.17)(10.369,7.000){2}{\rule{0.393pt}{0.400pt}}
\multiput(1387.00,804.59)(1.033,0.482){9}{\rule{0.900pt}{0.116pt}}
\multiput(1387.00,803.17)(10.132,6.000){2}{\rule{0.450pt}{0.400pt}}
\multiput(1399.00,810.59)(0.874,0.485){11}{\rule{0.786pt}{0.117pt}}
\multiput(1399.00,809.17)(10.369,7.000){2}{\rule{0.393pt}{0.400pt}}
\multiput(1411.00,817.59)(1.123,0.482){9}{\rule{0.967pt}{0.116pt}}
\multiput(1411.00,816.17)(10.994,6.000){2}{\rule{0.483pt}{0.400pt}}
\multiput(1424.00,823.59)(0.874,0.485){11}{\rule{0.786pt}{0.117pt}}
\multiput(1424.00,822.17)(10.369,7.000){2}{\rule{0.393pt}{0.400pt}}
\end{picture}
\end{center}
\end{figure}

\clearpage
\begin{figure}
\begin{center}
% GNUPLOT: LaTeX picture
\setlength{\unitlength}{0.240900pt}
\ifx\plotpoint\undefined\newsavebox{\plotpoint}\fi
\sbox{\plotpoint}{\rule[-0.200pt]{0.400pt}{0.400pt}}%
\begin{picture}(1500,900)(0,0)
\font\gnuplot=cmr10 at 10pt
\gnuplot
\sbox{\plotpoint}{\rule[-0.200pt]{0.400pt}{0.400pt}}%
\put(220.0,113.0){\rule[-0.200pt]{0.400pt}{184.048pt}}
\put(220.0,113.0){\rule[-0.200pt]{4.818pt}{0.400pt}}
\put(198,113){\makebox(0,0)[r]{{\large $0.48$}}}
\put(1416.0,113.0){\rule[-0.200pt]{4.818pt}{0.400pt}}
\put(220.0,304.0){\rule[-0.200pt]{4.818pt}{0.400pt}}
\put(198,304){\makebox(0,0)[r]{{\large $0.5$}}}
\put(1416.0,304.0){\rule[-0.200pt]{4.818pt}{0.400pt}}
\put(220.0,495.0){\rule[-0.200pt]{4.818pt}{0.400pt}}
\put(198,495){\makebox(0,0)[r]{{\large $0.52$}}}
\put(1416.0,495.0){\rule[-0.200pt]{4.818pt}{0.400pt}}
\put(220.0,686.0){\rule[-0.200pt]{4.818pt}{0.400pt}}
\put(198,686){\makebox(0,0)[r]{{\large $0.54$}}}
\put(1416.0,686.0){\rule[-0.200pt]{4.818pt}{0.400pt}}
\put(220.0,877.0){\rule[-0.200pt]{4.818pt}{0.400pt}}
\put(198,877){\makebox(0,0)[r]{{\large $0.56$}}}
\put(1416.0,877.0){\rule[-0.200pt]{4.818pt}{0.400pt}}
\put(220.0,113.0){\rule[-0.200pt]{0.400pt}{4.818pt}}
\put(220,68){\makebox(0,0){{\large $0$}}}
\put(220.0,857.0){\rule[-0.200pt]{0.400pt}{4.818pt}}
\put(410.0,113.0){\rule[-0.200pt]{0.400pt}{4.818pt}}
\put(410,68){\makebox(0,0){{\large $10$}}}
\put(410.0,857.0){\rule[-0.200pt]{0.400pt}{4.818pt}}
\put(600.0,113.0){\rule[-0.200pt]{0.400pt}{4.818pt}}
\put(600,68){\makebox(0,0){{\large $20$}}}
\put(600.0,857.0){\rule[-0.200pt]{0.400pt}{4.818pt}}
\put(790.0,113.0){\rule[-0.200pt]{0.400pt}{4.818pt}}
\put(790,68){\makebox(0,0){{\large $30$}}}
\put(790.0,857.0){\rule[-0.200pt]{0.400pt}{4.818pt}}
\put(980.0,113.0){\rule[-0.200pt]{0.400pt}{4.818pt}}
\put(980,68){\makebox(0,0){{\large $40$}}}
\put(980.0,857.0){\rule[-0.200pt]{0.400pt}{4.818pt}}
\put(1170.0,113.0){\rule[-0.200pt]{0.400pt}{4.818pt}}
\put(1170,68){\makebox(0,0){{\large $50$}}}
\put(1170.0,857.0){\rule[-0.200pt]{0.400pt}{4.818pt}}
\put(1360.0,113.0){\rule[-0.200pt]{0.400pt}{4.818pt}}
\put(1360,68){\makebox(0,0){{\large $60$}}}
\put(1360.0,857.0){\rule[-0.200pt]{0.400pt}{4.818pt}}
\put(220.0,113.0){\rule[-0.200pt]{292.934pt}{0.400pt}}
\put(1436.0,113.0){\rule[-0.200pt]{0.400pt}{184.048pt}}
\put(220.0,877.0){\rule[-0.200pt]{292.934pt}{0.400pt}}
\put(45,495){\makebox(0,0){{\Large $U$}}}
\put(828,-22){\makebox(0,0){{\Large $L$}}}
\put(828,-554){\makebox(0,0){{\large Figure 3}}}
\put(828,-650){\makebox(0,0){{\large Atsushi Yamagata}}}
\put(220.0,113.0){\rule[-0.200pt]{0.400pt}{184.048pt}}
\put(296,388){\circle{24}}
\put(334,358){\circle{24}}
\put(372,304){\circle{24}}
\put(410,270){\circle{24}}
\put(448,257){\circle{24}}
\put(486,254){\circle{24}}
\put(524,235){\circle{24}}
\put(562,240){\circle{24}}
\put(600,250){\circle{24}}
\put(638,272){\circle{24}}
\put(676,277){\circle{24}}
\put(714,292){\circle{24}}
\put(752,333){\circle{24}}
\put(790,317){\circle{24}}
\put(828,372){\circle{24}}
\put(866,412){\circle{24}}
\put(904,401){\circle{24}}
\put(942,423){\circle{24}}
\put(980,437){\circle{24}}
\put(1018,535){\circle{24}}
\put(1056,599){\circle{24}}
\put(1094,576){\circle{24}}
\put(1132,626){\circle{24}}
\put(1170,561){\circle{24}}
\put(1208,620){\circle{24}}
\put(1246,667){\circle{24}}
\put(1284,745){\circle{24}}
\put(1322,690){\circle{24}}
\put(1360,750){\circle{24}}
\put(296.0,386.0){\rule[-0.200pt]{0.400pt}{0.964pt}}
\put(286.0,386.0){\rule[-0.200pt]{4.818pt}{0.400pt}}
\put(286.0,390.0){\rule[-0.200pt]{4.818pt}{0.400pt}}
\put(334.0,356.0){\rule[-0.200pt]{0.400pt}{0.964pt}}
\put(324.0,356.0){\rule[-0.200pt]{4.818pt}{0.400pt}}
\put(324.0,360.0){\rule[-0.200pt]{4.818pt}{0.400pt}}
\put(372.0,301.0){\rule[-0.200pt]{0.400pt}{1.204pt}}
\put(362.0,301.0){\rule[-0.200pt]{4.818pt}{0.400pt}}
\put(362.0,306.0){\rule[-0.200pt]{4.818pt}{0.400pt}}
\put(410.0,267.0){\rule[-0.200pt]{0.400pt}{1.445pt}}
\put(400.0,267.0){\rule[-0.200pt]{4.818pt}{0.400pt}}
\put(400.0,273.0){\rule[-0.200pt]{4.818pt}{0.400pt}}
\put(448.0,253.0){\rule[-0.200pt]{0.400pt}{2.168pt}}
\put(438.0,253.0){\rule[-0.200pt]{4.818pt}{0.400pt}}
\put(438.0,262.0){\rule[-0.200pt]{4.818pt}{0.400pt}}
\put(486.0,250.0){\rule[-0.200pt]{0.400pt}{1.927pt}}
\put(476.0,250.0){\rule[-0.200pt]{4.818pt}{0.400pt}}
\put(476.0,258.0){\rule[-0.200pt]{4.818pt}{0.400pt}}
\put(524.0,230.0){\rule[-0.200pt]{0.400pt}{2.409pt}}
\put(514.0,230.0){\rule[-0.200pt]{4.818pt}{0.400pt}}
\put(514.0,240.0){\rule[-0.200pt]{4.818pt}{0.400pt}}
\put(562.0,236.0){\rule[-0.200pt]{0.400pt}{1.927pt}}
\put(552.0,236.0){\rule[-0.200pt]{4.818pt}{0.400pt}}
\put(552.0,244.0){\rule[-0.200pt]{4.818pt}{0.400pt}}
\put(600.0,243.0){\rule[-0.200pt]{0.400pt}{3.132pt}}
\put(590.0,243.0){\rule[-0.200pt]{4.818pt}{0.400pt}}
\put(590.0,256.0){\rule[-0.200pt]{4.818pt}{0.400pt}}
\put(638.0,266.0){\rule[-0.200pt]{0.400pt}{2.650pt}}
\put(628.0,266.0){\rule[-0.200pt]{4.818pt}{0.400pt}}
\put(628.0,277.0){\rule[-0.200pt]{4.818pt}{0.400pt}}
\put(676.0,270.0){\rule[-0.200pt]{0.400pt}{3.132pt}}
\put(666.0,270.0){\rule[-0.200pt]{4.818pt}{0.400pt}}
\put(666.0,283.0){\rule[-0.200pt]{4.818pt}{0.400pt}}
\put(714.0,284.0){\rule[-0.200pt]{0.400pt}{3.854pt}}
\put(704.0,284.0){\rule[-0.200pt]{4.818pt}{0.400pt}}
\put(704.0,300.0){\rule[-0.200pt]{4.818pt}{0.400pt}}
\put(752.0,327.0){\rule[-0.200pt]{0.400pt}{2.650pt}}
\put(742.0,327.0){\rule[-0.200pt]{4.818pt}{0.400pt}}
\put(742.0,338.0){\rule[-0.200pt]{4.818pt}{0.400pt}}
\put(790.0,290.0){\rule[-0.200pt]{0.400pt}{13.009pt}}
\put(780.0,290.0){\rule[-0.200pt]{4.818pt}{0.400pt}}
\put(780.0,344.0){\rule[-0.200pt]{4.818pt}{0.400pt}}
\put(828.0,349.0){\rule[-0.200pt]{0.400pt}{11.322pt}}
\put(818.0,349.0){\rule[-0.200pt]{4.818pt}{0.400pt}}
\put(818.0,396.0){\rule[-0.200pt]{4.818pt}{0.400pt}}
\put(866.0,382.0){\rule[-0.200pt]{0.400pt}{14.454pt}}
\put(856.0,382.0){\rule[-0.200pt]{4.818pt}{0.400pt}}
\put(856.0,442.0){\rule[-0.200pt]{4.818pt}{0.400pt}}
\put(904.0,374.0){\rule[-0.200pt]{0.400pt}{13.249pt}}
\put(894.0,374.0){\rule[-0.200pt]{4.818pt}{0.400pt}}
\put(894.0,429.0){\rule[-0.200pt]{4.818pt}{0.400pt}}
\put(942.0,385.0){\rule[-0.200pt]{0.400pt}{18.549pt}}
\put(932.0,385.0){\rule[-0.200pt]{4.818pt}{0.400pt}}
\put(932.0,462.0){\rule[-0.200pt]{4.818pt}{0.400pt}}
\put(980.0,401.0){\rule[-0.200pt]{0.400pt}{17.345pt}}
\put(970.0,401.0){\rule[-0.200pt]{4.818pt}{0.400pt}}
\put(970.0,473.0){\rule[-0.200pt]{4.818pt}{0.400pt}}
\put(1018.0,512.0){\rule[-0.200pt]{0.400pt}{11.081pt}}
\put(1008.0,512.0){\rule[-0.200pt]{4.818pt}{0.400pt}}
\put(1008.0,558.0){\rule[-0.200pt]{4.818pt}{0.400pt}}
\put(1056.0,577.0){\rule[-0.200pt]{0.400pt}{10.600pt}}
\put(1046.0,577.0){\rule[-0.200pt]{4.818pt}{0.400pt}}
\put(1046.0,621.0){\rule[-0.200pt]{4.818pt}{0.400pt}}
\put(1094.0,539.0){\rule[-0.200pt]{0.400pt}{18.067pt}}
\put(1084.0,539.0){\rule[-0.200pt]{4.818pt}{0.400pt}}
\put(1084.0,614.0){\rule[-0.200pt]{4.818pt}{0.400pt}}
\put(1132.0,591.0){\rule[-0.200pt]{0.400pt}{16.863pt}}
\put(1122.0,591.0){\rule[-0.200pt]{4.818pt}{0.400pt}}
\put(1122.0,661.0){\rule[-0.200pt]{4.818pt}{0.400pt}}
\put(1170.0,521.0){\rule[-0.200pt]{0.400pt}{19.272pt}}
\put(1160.0,521.0){\rule[-0.200pt]{4.818pt}{0.400pt}}
\put(1160.0,601.0){\rule[-0.200pt]{4.818pt}{0.400pt}}
\put(1208.0,568.0){\rule[-0.200pt]{0.400pt}{25.294pt}}
\put(1198.0,568.0){\rule[-0.200pt]{4.818pt}{0.400pt}}
\put(1198.0,673.0){\rule[-0.200pt]{4.818pt}{0.400pt}}
\put(1246.0,625.0){\rule[-0.200pt]{0.400pt}{20.476pt}}
\put(1236.0,625.0){\rule[-0.200pt]{4.818pt}{0.400pt}}
\put(1236.0,710.0){\rule[-0.200pt]{4.818pt}{0.400pt}}
\put(1284.0,712.0){\rule[-0.200pt]{0.400pt}{15.899pt}}
\put(1274.0,712.0){\rule[-0.200pt]{4.818pt}{0.400pt}}
\put(1274.0,778.0){\rule[-0.200pt]{4.818pt}{0.400pt}}
\put(1322.0,654.0){\rule[-0.200pt]{0.400pt}{17.586pt}}
\put(1312.0,654.0){\rule[-0.200pt]{4.818pt}{0.400pt}}
\put(1312.0,727.0){\rule[-0.200pt]{4.818pt}{0.400pt}}
\put(1360.0,711.0){\rule[-0.200pt]{0.400pt}{19.031pt}}
\put(1350.0,711.0){\rule[-0.200pt]{4.818pt}{0.400pt}}
\put(1350.0,790.0){\rule[-0.200pt]{4.818pt}{0.400pt}}
\end{picture}
\end{center}
\end{figure}
\end{document}